# Towards More Flexible False Positive Control in Phase III Randomized Clinical Trials

**Changyu Shen[1,*] and Xiaochun Li[2]**

[1]Beth Israel Deaconess Medical Center, Harvard Medical School, Boston, USA
[2]Department of Biostatistics, Indiana University School of Medicine, Indianapolis, USA

**Abstract**

Phase III randomized clinical trials play a monumentally critical role in the evaluation of new medical products. Because of the intrinsic nature of uncertainty embedded in our capability in assessing the efficacy of a medical product, interpretation of trial results relies on statistical principles to control the error of false positives below desirable level. The well-established statistical hypothesis testing procedure suffers from two major limitations, namely, the lack of flexibility in the thresholds to claim success and the lack of capability of controlling the total number of false positives that could be yielded by the large volume of trials. We propose two general theoretical frameworks based on the conventional frequentist paradigm and Bayesian perspectives, which offer realistic, flexible and effective solutions to these limitations. Our methods are based on the distribution of the effect sizes of the population of trials of interest. The estimation of this distribution is practically feasible as clinicaltrials.gov provides a centralized data repository with unbiased coverage of clinical trials. We provide a detailed development of the two frameworks with numerical results obtained for industry sponsored Phase III randomized clinical trials.

*Address of correspondence*: Changyu Shen, Smith Center for Outcomes Research in Cardiology, Beth Israel Deaconess Medical Center, Harvard Medical School, 375 Longwood Avenue, Boston, Massachusetts 02215, USA. E-mail: cshen1@bidmc.harvard.edu

# 1 Introduction

The conventional statistical hypothesis testing procedure has been widely adopted in biomedical research as the foundational framework for study designs and data analyses. Nonetheless, there have been continuous criticisms on the $p<0.05$ criterion as the primary driving force of the conclusion of a study. Issues raised in these criticisms include (a) confusion among investigators between statistical significance and clinical significance; (b) the incorrect interpretation of lack of evidence of effect being equal to evidence of lack of effect for $p \geq 0.05$ [1]; (c) false interpretation of *p*-value being the probability that the null hypothesis is true [2]; (d) the universally applied threshold of 0.05 may not be appropriate for all situations [3,4]. Benjamin et al. proposed to make the statistical significance criterion more stringent by changing the threshold from 0.05 to 0.005 to improve reproducibility [3]. Nevertheless, such a change may cause more confusions in the research community on issues (a) and (b). McShane et al. proposed an alternative solution by demoting *p*-value from the predominate piece of evidence to one of many pieces of evidences [4]. Nevertheless, the authors did not provide specifics on how errors can be controlled or estimated by combining different sources of evidence. For example, as conclusions or decision-makings are often intrinsically discrete, some thresholds (one way or another) are needed. Yet, there is no discussion on how thresholds should be chosen in McShane et al [4].

Statistical inference of the efficacy of a medical intervention is of paramount importance in Phase III randomized clinical trials (RCTs), which represent perhaps the most rigorous clinical investigations by any standard with tremendous financial cost. The aforementioned issues with *p*-values are of great relevance to how data generated from the huge endeavors devoted to RCTs should be interpreted, which has significant public health impact as Phase III RCTs represent the last layer of scrutiny before a medical product is put on market to affect a large number of patients.

Apparently, the threshold for the *p*-value plays a major role here. In spirit, both Benjamin et al. [3] and McShane et al. [4] support the idea that the threshold for *p*-values should be made different from studies to studies to accommodate the specific context of each study. We agree with this view and argue that a universally applied threshold is unlikely to work well for all Phase III RCTs. There is heterogeneity at different levels among Phase III RCTs. There are scientific, practical and ethic factors that could affect how we want to approach the uncertainty of the efficacy of a medical intervention. For example, if early phase research demonstrates strong potential in treatment efficacy, relaxing the threshold of the *p*-value for a Phase III trial is justified from the perspectives of cost (relaxed threshold means smaller sample size) and ethics (smaller sample size means less patients in the control arm). On the other hand, if a drug has huge number of potential consumers (so that false positive has significant impact) and the earlier phase result is not particularly strong, then a more stringent threshold may be desirable.

In this article, we propose two statistical frameworks that allow for flexible false positive control for each individual Phase III RCT and in the meanwhile control for the expected number of false positives (ENFPs) yielded from the totality of trials in a given period. The first framework is based on the currently well-adopted hypothesis testing procedure. The second framework is built on a Bayes perspective. We believe these methods are effective solutions to issue (d) discussed at the beginning of this section. The second framework has two extra advantages. First, it provides a direct assessment on the magnitude of the efficacy with a continuous scale of uncertainty through the posterior probability, as opposed to the conventional *p*-value that cannot be directly interpreted as an uncertainty measure on the magnitude of efficacy. This advantage effectively avoids issues (a)-(c). As this article is focused on hypothesis testing, we refer interested readers to Shen [5] for more details on this aspect. Second, as the posterior probability of null hypothesis being true is a monotone function of the Z statistic [6] and consequently a monotone function of the *p*-value, the threshold can be executed at the scale of the posterior probability, which can be more interpretable.

Central to our strategy is the distribution of the effect size of recently completed Phase III RCTs as a prior distribution. The focus on effect size and Bayesian perspective are also proposed recently as solutions for the improvement of statistical inference [7]. In a previous study, we have demonstrated how this distribution can be estimated using trials registered at clinicaltrials.gov (CT.gov) [8]. In what follows, we introduce the notations and describe the first framework based on the widely adopted statistical hypothesis testing paradigm. We then describe the second framework from the Bayes perspective. Lastly, we provide some numerical results and conclude the article with a discussion.

## 2 Method based on frequentist hypothesis testing

### 2.1 Notations

The setup of our framework is similar to a random-effects ANOVA or meta-analysis. We will first introduce notation of parameters and statistics for a fixed trial, followed by the extension to a population of trials where parameters are considered random. Bold symbols refer to vectors. We will use the term *intervention* to refer to a new medical product being evaluated and *control* to refer to a suitable comparison treatment.

For a given trial, we will use $\beta$ to denote an *efficacy measure*, which is an unknown numerical quantity summarizing the efficacy of the intervention relative to the control for a specific endpoint, e.g. $\beta$ can be the logarithm of the odds ratio of mortality of the control arm over the intervention arm. As a convention, we assume higher value of $\beta$ indicates higher efficacy of the intervention. As some Phase III RCTs have multiple endpoints and/or multiple comparison groups, we will use $\boldsymbol{\beta} = \{\beta_i, i = 1,2 \dots m\}$ to denote the vector of efficacy measures of a trial. For each trial, the *failure*

*region* ($\mathcal{F}$) is the set of $\boldsymbol{\beta}$ values corresponding to the definition of null efficacy. There are in general two types of failure regions:

$$\text{Type A: } \mathcal{F} = \cap_{i=1}^{m}\{\beta_i \leq c_i\}$$

$$\text{Type B: } \mathcal{F} = \cup_{i=1}^{m}\{\beta_i \leq c_i\},$$

where $c_i$ is the threshold deemed appropriate for efficacy measure $i$. The two types of failure regions cover all typical testing scenarios including superiority trials, non-inferiority trials, equivalence trials and their combinations. Note that equivalence trials can be viewed as having type B failure region with two $\beta$'s defined the same way except the positions of the two comparison arms are switched. By convention, we will call the failure region of a trial with $m = 1$ a type B failure region. A trial is a *false positive* if we claim $\boldsymbol{\beta} \in \mathcal{F}^c$ when in fact $\boldsymbol{\beta} \in \mathcal{F}$. Let $\boldsymbol{D}$ be the vector of test statistics for the $m$ comparisons. $\mathcal{R}$ is the *rejection region* in the sample space of $\boldsymbol{D}$ such that we claim $\boldsymbol{\beta} \in \mathcal{F}^c$ if $\boldsymbol{D} \in \mathcal{R}$. The *type I error rate* is $\alpha = Max_{\boldsymbol{\beta}^* \in \mathcal{F}} \Pr[\boldsymbol{D} \in \mathcal{R}|\boldsymbol{\beta}^*]$. Control of $\alpha$ is through the specification of $\mathcal{R}$, which is adjusted through multiple-comparison procedure for $m > 1$. Note that $(\mathcal{F}, \mathcal{R}, \alpha)$ is determined by the investigator(s) of a trial.

Suppose there is a well-defined population of trials, where the rejection regions are such that the type I error rates are allowed to be different from trials to trials. This population of trials induce a distribution of $(\boldsymbol{\beta}, \mathcal{F}, \mathcal{R}, \alpha)$, where $(\boldsymbol{\beta}, \mathcal{F}, \mathcal{R}, \alpha)$ is considered random. For the sake of argument, we will view this distribution as an empirical distribution with a large finite number of quadruplets (and therefore sufficient granularity) to avoid technicality in defining a random vector with variable dimension ($\boldsymbol{\beta}$) and a random set ($\mathcal{F}$ and $\mathcal{R}$). We assume the index of the efficacy measures $i$ is not informative of the value of the efficacy measure. In other words, $L(\beta_i) = L(\beta_j), i \neq j$, where $L(X)$ is the law of $X$. Let $\pi = \Pr[\boldsymbol{\beta} \in \mathcal{F}]$ be the proportion of trials with null efficacy. Similarly, $\rho = \Pr[\beta \leq c]$, where $(\beta, c)$ is a randomly selected pair of efficacy measure and the corresponding threshold from the pool of pairs generated by the entire population of trials. The expected number of false positives (ENFP) for $N$ trials is defined as

$$\text{ENFP} = N\Pr[\boldsymbol{D} \in \mathcal{R}, \boldsymbol{\beta} \in \mathcal{F}]. \tag{1}$$

**2.2 Flexible type I error rate with control on ENFP ($m=1$)**

When $m = 1$ for all trials, we can drop the subscript for $\beta$ and $c$,

$$\Pr[\boldsymbol{D} \in \mathcal{R}, \beta \leq c] = \Pr[\beta \leq c]\Pr[\boldsymbol{D} \in \mathcal{R}|\beta \leq c] \leq \Pr[\beta \leq c]E[\alpha|\beta \leq c] = \rho E[\alpha|\beta \leq c]. \tag{2}$$

We make the following assumption:

*First Concordance Assumption:* $E[\alpha|\beta \leq c] \leq E[\alpha|\beta > c]$: the type I error rate control for those without efficacy is on average at least as stringent as those with efficacy.

The first concordance assumption implies that

$$E[\alpha] = \rho E[\alpha|\beta \leq c] + (1-\rho)E[\alpha|\beta > c] \geq \rho E[\alpha|\beta \leq c] + (1-\rho)E[\alpha|\beta \leq c]$$

$$= E[\alpha|\beta \leq c].$$

Combining equation (2) and the inequality above, we have

$$\text{ENFP} = N\Pr[\boldsymbol{D} \in \mathcal{R}, \boldsymbol{\beta} \in \mathcal{F}] \leq N\rho E[\alpha|\beta \leq c] \leq N\rho E[\alpha] = \tau. \tag{3}$$

With an estimate of $\rho$ (estimation of $\rho$ is discussed in a later section), $\hat{\rho}$, we can estimate the upper bound $\tau$ by

$$\hat{\tau} = \hat{\rho} \sum_{i=1}^{N} \alpha_i, \tag{4}$$

where $\alpha_i$ is the type I error rate for trial $i$.

Therefore, we can allow different trials to have different type I error rates and are able to estimate an upper bound of the ENFP. The only requirement is the first concordance assumption, that is, we assign to those trials without efficacy a type I error rate that is on average at least as stringent as that of the trials with efficacy. This assumption is reasonable in practical settings. A natural strategy is to relax the threshold for high likelihood of efficacy and tighten the threshold for low likelihood of efficacy. The existing evidence on efficacy before a trial starts could come from earlier phase study results, knowledge of biological mechanism or efficacy of similar products. Although the existing evidence has uncertainty, it is extremely unlikely that the first concordant assumption will be violated under this practice. The worst that could happen is perhaps there is no difference on average in the type I error rate between those trials with and without efficacy, in which case the first concordance assumption still holds.

## 2.3 Flexible type I error rate with control on ENFP (*m*>1 for some trials)

A more realistic scenario is that the population of trials is composed of trials with variable $m$. With the same derivation as in (2), we have

$$\Pr[\boldsymbol{D} \in \mathcal{R}, \boldsymbol{\beta} \in \mathcal{F}] \leq \Pr[\boldsymbol{\beta} \in \mathcal{F}]E[\alpha|\boldsymbol{\beta} \in \mathcal{F}] = \pi E[\alpha|\boldsymbol{\beta} \in \mathcal{F}]. \tag{5}$$

With the same spirit as in the first concordance assumption, we can make the following assumption:

<u>Second Concordance Assumption:</u> $E[\alpha|\boldsymbol{\beta} \in \mathcal{F}] \leq E[\alpha|\boldsymbol{\beta} \in \mathcal{F}^c]$: the type I error rate for those without overall efficacy is at least as stringent as those with overall efficacy.

With the second concordant assumption, we have

$$\text{ENFP} \leq N\pi E[\alpha]. \tag{6}$$

Let $t$ indicate the type of a failure region (i.e. $t = A, B$) and $\pi(m, t) = \Pr[\boldsymbol{\beta} \in \mathcal{F}|m, t]$. For simplicity, we will assume $\Pr[\beta \leq c|m, t] = \Pr[\beta \leq c] = \rho$. In other words, the probability of null efficacy does not depend on the type of failure region and number of efficacy measures. If this does not hold, then an analysis stratified by $t$ and/or $m$ can be performed (see last paragraph of Section "Implementation"). Then

$$\pi(m, t = A) \leq \rho \text{ and } \pi(m, t = B) \leq m\rho.$$

Let $\delta(\rho, m, t = A) = \rho$ and $\delta(\rho, m, t = B) = m\rho$. Then

$$\pi = E[\pi(m, t)] \leq E[\delta(\rho, m, t)], \tag{7}$$

where the expectation is with respect to the distribution of $(m, t)$. Combining (6) and (7) leads to

$$\text{ENFP} \leq \tau = NE[\delta(\rho, m, t)]E[\alpha].$$

With an estimate $\hat{\rho}$, we can estimate $\tau$ by

$$\hat{\tau} = \frac{1}{N}\left(\sum_{i=1}^{N} \delta(\hat{\rho}, m_i, t_i)\right)\left(\sum_{i=1}^{N} \alpha_i\right), \tag{8}$$

where $m_i$ and $t_i$ are the number of efficacy measures and type of failure region for trial $i$.

## 2.4 Implementation

Equations (4) and (8) are the basis for the control of ENFP as a global false positive measure while in the meanwhile allowing for flexible type I error rate control of individual trials. We first obtain $\hat{\rho}$ from recent historical data. From a regulatory perspective, we can choose a desirable threshold $\tau_0$ within a period of time (e.g. next five years) and set $\hat{\tau}$ to 0. With each additional trial, $\hat{\tau}$ will be updated accumulatively based on equations (4) and (8) until it reaches $\tau_0$. In this process, we can decide $\alpha_i$ for trial $i$ based on the specific context as long as the process that generates $\alpha_i$ does not violate the concordance assumptions. Note that this process also does not require the knowledge of $N$ at the beginning. Apparently, the process of randomly selecting $\alpha_i$ from some continuous or discrete regions in $(0,1)$ or fix $\alpha_i$ (e.g. current practice) meets the concordance assumptions.

When $m = 1$ for all trials, ENFP control can be achieved by requiring $\sum_{i=1}^{N} \alpha_i \leq \tau_0/\hat{\rho}$. In other words, $\tau_0/\hat{\rho}$ can be viewed as the *total error* that can be spent on trials during the specific period of time in a possibly non-uniform manner, e.g. an alpha spending strategy.

It should be noted that the strategy just described can be implemented in a stratified fashion. For example, we can divide trials into different categories according to the nature of the trials (e.g. types of diseases, interventions, and efficacy measurements etc.). Within each stratum, we can implement the strategy just described with a stratum-specific $\rho$. Similarly, we can also divide trials based on $t$ and/or $m$ if there is evidence that $\Pr[\beta \leq c|m, t] \neq \rho$ for at least some $t$ and/or $m$.

Within each stratum, we can control for ENFP and allow flexible type I error control for individual trials as long as we can estimate $\rho(t, m) = \Pr[\beta \leq c | m, t]$ within each stratum.

## 3 Method based on Bayesian perspective

### 3.1 Flexible trial-specific threshold with control on ENFP (*m*=1)

The treatment effect $\beta$ can be consistently estimated by $\hat{\beta}$ using well-established methods, which has a normal distribution (or at least approximately so) in most Phase III trials where sample sizes are relatively large. Let $\sigma$ denote the standard error of $\hat{\beta}$. Usually $\sigma$ is unknown. Nonetheless, $\sigma$ can be replaced by the standard estimator without impact on the inference procedure to be discussed in this section [5]. Therefore, without loss of generality, we will assume $\sigma$ is known. Let $\theta = (\beta - c)/\sigma$ so that positive efficacy is equivalent to $\theta > 0$. The conventional Z statistic is $Z = (\hat{\beta} - c)/\sigma$, which is (approximately) normally distributed with mean $\theta$ and standard deviation 1. Therefore, $Z$ can be viewed as a noisy version of $\theta$. The vector $\boldsymbol{D}$ essentially includes the $Z$ values for the test of each efficacy measure. Apparently, when $m = 1$, $\boldsymbol{D} = Z$. But we will continue using the notation $\boldsymbol{D} \in \mathcal{R}$ with the understanding that it means $Z = z$ is greater than the critical value corresponding to $\mathcal{R}$.

If we know the distribution of $\theta$ across the population of trials, $g(\theta)$, we can compute the posterior probability of positive efficacy (i.e. $\theta > 0$) for a given efficacy measure with $Z = z$ as

$$h(z) = \Pr[\theta > 0 | Z = z] = \int I(\theta > 0) g(\theta) \phi(z - \theta) d\theta \Big/ \int g(\theta) \phi(z - \theta) d\theta, \qquad (9)$$

where $\phi(\cdot)$ is the probability density function of the standard normal distribution. $h(z)$ has been termed as the *h*-probability to emphasize that $g(\theta)$ is based on results of historical trials [6]. It captures the proportion of trials with positive efficacy among all trials with $Z = z$. The *h*-probability can also be viewed as 1 minus the local false discovery rate (FDR) [9]. It has been shown that $h(z)$ is a monotone increasing function of $z$ [6] so that rejection regions defined by the *p*-value,

$z$ and $h(z)$ are all equivalent. Note that $h(z)$ has a unique advantage over the other two measures as its interpretation is more sensible, which makes it heuristically a natural choice for the determination of the threshold.

Shen and Li proposed an assumption that allows the use of $h(z)$ for the estimation of ENFP [6]. The assumption can be summarized as the following:

$$\Pr[\boldsymbol{d} \in \mathcal{R}|\theta_1] \leq \Pr[\boldsymbol{d} \in \mathcal{R}|\theta_2] \text{ for any } \theta_1 \leq 0, \theta_2 > 0 \text{ and fixed statistic } \boldsymbol{d}. \quad (10)$$

Equation (10) essentially means the rejection region is stochastically smaller for any given trial with null efficacy as compared with any given trial with positive efficacy. Assumption (10) implies the first concordance assumption but not necessarily vice versa. To see this, note inequality (10) means $S(\cdot|\theta_1) \leq S(\cdot|\theta_2)$, where $S(\cdot|\theta)$ is the survival function of $\alpha$ given $\theta$. Because $E(\alpha|\theta) = \int S(\alpha|\theta) \, d\alpha$, it follows that $E(\alpha|\theta_1) \leq E(\alpha|\theta_2)$ and then $E(\alpha|\theta \leq 0) \leq E(\alpha|\theta > 0)$.

Under (10), it can be shown that $\Pr[\theta \leq 0|Z = z, \boldsymbol{D} \in \mathcal{R}] \leq \Pr[\theta \leq 0|Z = z] = 1 - h(z)$ [6]. Then we have

$$\text{ENFP} = N\Pr[\boldsymbol{D} \in \mathcal{R}]E\{\Pr[\theta \leq 0|Z, \boldsymbol{D} \in \mathcal{R}] \,|\boldsymbol{D} \in \mathcal{R}\} \leq ME\{1 - h(Z)|\boldsymbol{D} \in \mathcal{R}\} = \omega,$$

where $M = N\Pr[\boldsymbol{D} \in \mathcal{R}]$ is the expected number of positive trials. Then we can use the following estimator to estimate the upper bound of ENFP

$$\widehat{\omega} = \sum_{i \in P}(1 - \hat{h}(z_i)), \quad (11)$$

where $P$ is the set of positive trials, $z_i$ is the observed Z value for trial $i \in P$, and $\hat{h}(z_i)$ is computed by plugging an estimate $\hat{g}(\theta)$ in equation (9). The advantage of (11) over (4) is that estimator in (11) can be closer to ENFP than (4) because it accounts for the actual negative value of $\theta$ instead of assuming $\theta \equiv 0$ for trials with null efficacy.

Shen and Li [6] also pointed out that the definition of ENFP in equation (1) averages over all possible values of $\boldsymbol{Z}$ that is composed of the $N$ $Z$ statistics. Nonetheless, we only observe one realization of these Z values in reality. Thus, the definition of ENFP in equation (1) may not be very "relevant". We may want to consider an alternative definition of ENFP that is conditional on the observed $z$ in the set of positive trials. A natural way to quantify the propensity of the intervention being a false positive is $\Pr[\theta \leq 0 | Z = z, \mathbf{D} \in \mathcal{R}]$. The computation of $\Pr[\theta \leq 0 | Z = z, \mathbf{D} \in \mathcal{R}]$ requires knowledge of $g(\theta | \mathbf{D} \in \mathcal{R})$, which is more difficult to estimate than $g(\theta)$ for two reasons. First, there are fewer historical positive trials compared with the whole collection of historical trials, reducing estimation precision. Second, the rejection region can be constantly changing once the flexible error control strategy is implemented. Thus, the estimate from history may not reflect the presence. As an alternative, we can use $1 - h(z)$ as a measure of propensity of false positive. Along this line, we propose a third concordance assumption that is weaker than (10) (e.g. the third concordance assumption holds if (10) is true but not necessarily vice versa [6]).

*Third Concordance Assumption:* $\Pr[\theta \leq 0 | Z = z, \mathbf{D} \in \mathcal{R}] \leq \Pr[\theta \leq 0 | Z = z, \mathbf{D} \in \mathcal{R}^c]$ for all $z$ such that there is non-zero density on the set $\{Z = z, \mathbf{D} \in \mathcal{R}\}$ and $\{Z = z, \mathbf{D} \in \mathcal{R}^c\}$, the probability of null efficacy among positive trials is no more than that of the negative trials within trials with $Z = z$.

Under the third concordance assumption, we have

$$\Pr[\theta \leq 0 | Z = z, \mathbf{D} \in \mathcal{R}] \leq \Pr[\theta \leq 0 | Z = z] = 1 - h(z). \tag{12}$$

Therefore, we can use $1 - h(z)$ as a more conservative quantification of the propensity of null efficacy given $Z = z$. Below we define an alternative version of ENFP by fixing the observed Z values in the set $P$ that is composed of the positive trials, i.e. $\mathbf{z}_P = \{z_i : i \in P\}$:

$$\text{ENFP}_{cond}(\mathbf{z}_P) = \sum_{i \in P} \Pr[\theta_i \leq 0 | Z_i = z_i] = \sum_{i \in P}(1 - h(z_i)). \tag{13}$$

Inequality (12) implies

$$E(1 - h(Z_i)|\boldsymbol{D}_i \in \mathcal{R}_i) \geq \boldsymbol{E}\{\Pr[\theta_i \leq 0|Z_i = z_i, \boldsymbol{D}_i \in \mathcal{R}_i]|\boldsymbol{D}_i \in \mathcal{R}_i\} = \Pr[\theta_i \leq 0|\boldsymbol{D}_i \in \mathcal{R}_i],$$

which means the definition of $\text{ENFP}_{cond}(\boldsymbol{z}_P)$ in (13) is conservative. The estimator in (11) can be directly used for the estimation of $\text{ENFP}_{cond}(\boldsymbol{z}_P)$.

The execution of the flexible error control from the Bayes perspective is similar to what is described previously for the frequentist method. For a trial, we can define $\mathcal{R}$ based on early phase efficacy evidence and other factors related to efficacy assessment, e.g. less stringent $\mathcal{R}$ for interventions with stronger evidence. Once data are collected we add $1 - \hat{h}(z)$ to $\widehat{\omega}$ in (11) if $\boldsymbol{D} \in \mathcal{R}$. Therefore, for trials with the same $z$, those deemed positive are the ones with a relaxed $\mathcal{R}$ as the result of some evidence of efficacy from other perspectives. According to the third concordance assumption, these trials should be at least not more likely to have null efficacy as compared with the ones deemed negative (because of a stringent $\mathcal{R}$) that have less evidence of efficacy from other perspectives. Certainly, strong early phase result could be generated by large error, in which case the actual efficacy may not be that strong (or regression to the mean). Nevertheless, it is extremely unlikely that using this kind of evidence to determine rejection region will lead to a violation of the third concordance assumption. Trivial cases are very small or large $z$, where all trials with that value will be deemed negative or positive.

### 3.2 Flexible trial-specific threshold with control on ENFP (*m*>1 for some trials)

Suppose $\theta^{(j)}$ and $Z^{(j)}$ ($j = 1,2 \ldots, m$) are the effect size and $Z$ statistic corresponding to the *j*th efficacy measure. For type A failure region, the event "$\theta^{(1)} \leq 0$" has equal or higher probability than the event $\boldsymbol{\beta} \in \mathcal{F}$. Thus, we can consider $\Pr[\theta^{(1)} \leq 0|Z^{(1)} = z^{(1)}, \boldsymbol{D} \in \mathcal{R}, t = A, m]$ as a conservative measure of the propensity of null efficacy for a positive trial. For type B failure region, we can consider $\sum_{j=1}^{m} \Pr[\theta^{(j)} \leq 0|Z^{(j)} = z^{(j)}, \boldsymbol{D} \in \mathcal{R}, t = B, m]$. We make the following assumption:

*Fourth Concordance Assumption:*

$\Pr[\theta^{(j)} \leq 0 | Z^{(j)} = z, \boldsymbol{D} \in \mathcal{R}, t, m] \leq \Pr[\theta^{(j)} \leq 0 | Z^{(j)} = z, \boldsymbol{D} \in \mathcal{R}^c, t, m]$ for all $t, m, 1 \leq j \leq m$ and $z$ such that there is non-zero density on the set $\{Z^{(j)} = z, \boldsymbol{D} \in \mathcal{R}, t, m\}$, and $\{Z^{(j)} = z, \boldsymbol{D} \in \mathcal{R}^c, t, m\}$: for all $t, m, 1 \leq j \leq m$ and $z$, the probability of null efficacy for the $j$th efficacy measure from positive trials is no more than that from the negative trials among trials with failure region $t$, $m$ efficacy measures and $Z^{(j)} = z$.

Under the fourth concordance assumption, we have

$$\Pr[\theta^{(j)} \leq 0 | Z^{(j)} = z, \boldsymbol{D} \in \mathcal{R}, t, m] \leq \Pr[\theta^{(j)} \leq 0 | Z^{(j)} = z, t, m].$$

Thus, we can use $\Pr[\theta^{(1)} \leq 0 | Z^{(1)} = z^{(1)}, t = A, m]$ or $\sum_{j=1}^{m} \Pr[\theta^{(j)} \leq 0 | Z^{(j)} = z^{(j)}, t = B, m]$ to conservatively quantify the propensity of null efficacy for a positive trial with type A or B failure regions. Similar to the counterpart in frequentist hypothesis testing, we will assume the distribution of the effect size does not depend on $t$ and/or $m$, i.e. $g(\theta|m, t) = g(\theta)$. Under this assumption, the posterior distribution of $\theta$ does not depend on $t$ and $m$. Therefore, we can further simplify the propensity measure to $\Pr[\theta^{(1)} \leq 0 | Z^{(1)} = z^{(1)}]$ or $\sum_{j=1}^{m} \Pr[\theta^{(j)} \leq 0 | Z^{(j)} = z^{(j)}]$. We can then define $\text{ENFP}_{cond}(\boldsymbol{z}_P)$ as

$\text{ENFP}_{cond}(\boldsymbol{z}_P)$

$$= \sum_{i \in P} \left\{ I(t_i = A) \Pr\left[\theta_i^{(1)} \leq 0 \middle| Z_i^{(1)} = z_i^{(1)}\right] + I(t_i = B) \sum_{j=1}^{m_i} \Pr\left[\theta_i^{(j)} \leq 0 \middle| Z_i^{(j)} = z_i^{(j)}\right] \right\}$$

$$= \sum_{i \in P} \left\{ I(t_i = A) \left[1 - h\left(z_i^{(1)}\right)\right] + I(t_i = B) \sum_{j=1}^{m_i} \left[1 - h\left(z_i^{(j)}\right)\right] \right\}$$

$$= \sum_{i \in P} G_i, \qquad (14)$$

where $z_i^{(j)}$ is the observed Z statistic for the jth efficacy measure of positive trial i. Similar to the scenario of m=1, it can be shown that $E(G_i| \boldsymbol{D_i} \in \mathcal{R}_i) \geq \Pr[\boldsymbol{\beta_i} \in \mathcal{F}_i|\boldsymbol{D_i} \in \mathcal{R}_i]$ so that the $\text{ENFP}_{cond}(\boldsymbol{z}_P)$ in (14) is conservative. Similarly to the $m = 1$ scenario, $\text{ENFP}_{cond}(\boldsymbol{z}_P)$ can be estimated by

$$\hat{\omega} = \sum_{i \in P} \left\{ I(t_i = \text{A}) \left[1 - \hat{h}\left(z_i^{(1)}\right)\right] + I(t_i = \text{B}) \sum_{j=1}^{m_i} \left[1 - \hat{h}\left(z_i^{(j)}\right)\right] \right\}. \tag{15}$$

Again, if $g(\theta|m,t) \neq g(\theta)$, stratified analysis can be performed separately for each stratum defined by $t$ and/or $m$.

### 3.3 Implementation

Equations (11) and (15) are the basis for the implementation of the flexible error control from the Bayes perspective. At the beginning of a time period, $\hat{\omega}$ is set to 0. For each trial, the rejection region can be defined by applying a threshold to the h-probability (equation (9)) for each efficacy measure, which can be made different from trials to trials. With each additional positive trial, $\hat{\omega}$ will be updated accumulatively based on equations (11) and (15) until it reaches a pre-specified threshold $\omega_0$. One advantage of the Bayes method is that it allows for potential adjustment after data are collected. For example, suppose there is a costly trial on a target disease population where enrollment is very difficult. The trial only reached 75% of the targeted enrollment with a p-value short of the pre-specified threshold (e.g. threshold is 0.05 but we got 0.09). There is clearly a possibility that the insignificance is caused by insufficient sample size. Unfortunately, the conventional frequentist method is incapable of making any adjustment in the decision process to address the sample size issue. This is because the alpha has to be determined before data are collected. On the other hand, the Bayes method allows us to compute the contribution to the number of false positive (e.g. 1 minus h-probability) if we decide to claim it is a positive trial. Certainly, such an adjustment could raise concern on the validity of the third or fourth concordance assumption as it could "contaminate" the distribution of $\theta$ in the positive trials (e.g. shifting the distribution to the left if the intervention has null efficacy). Nonetheless, as long as the frequency of such an adjustment is small, it is unlikely to reverse the relationship of $E(G_i|\boldsymbol{D_i} \in \mathcal{R}_i) \geq \Pr[\boldsymbol{\beta_i} \in \mathcal{F}_i|\boldsymbol{D_i} \in \mathcal{R}_i]$ and the estimators (11) and (15) are still conservative.

## 4 Estimation of Parameters

The execution strategies described in previous sections requires estimates of $\rho = \Pr[\theta \leq 0]$ and $g(\theta)$. Because $\rho = \int I(\theta \leq 0)g(\theta)d\theta$, we can obtain $\hat{\rho} = \int I(\theta \leq 0)\hat{g}(\theta)d\theta$ once we obtain $\hat{g}(\theta)$. If we can collect a representative sample of $Z$ values from the underlying population of trials, we can then apply deconvolution method to compute $\hat{g}(\theta)$. There is a rich literature on parametric, semi-parametric and non-parametric deconvolution methods [10-12]. We consider a semi-parametric method developed by Efron that strikes a balance between robustness and efficiency [11].

Under the Food and Drug Administration Amendments Act (FDAAA), all Phase III clinical trials of drugs, medical devices and biologics that were initiated after September 27, 2007, or were ongoing as of December 26, 2007 [13], are required to be registered at clinicaltrials.gov (CT.gov) regardless of trial results. Trials on medical products approved by the Food and Drug Administration (FDA) are also required to submit aggregated trial results. This law provides an unprecedented opportunity to identify the entire population of trials and their results, which makes the computation of $\hat{g}(\theta)$ realistic. In a previous study, we identified all Phase III randomized superiority trials completed in 2008-2012 that were at least partially sponsored by industry under the oversight of the FDA [8]. Among the 1393 trials that met our entry criteria, we selected one efficacy measure from each trial and sought to identify the *p*-values through a variety of resources including CT.gov, publications, conference abstracts, press release, clinical summary report from the sponsor and direct contact of the trial sponsor [8]. The reason to resort to other resources beyond

CT.gov is that results of trials on medical products not yet approved by the FDA may not be reported at CT.gov or results on FDA approved products are not stored at CT.gov due to non-compliance [14, 15]. We were able to retrieve information on the *p*-values for 1221 efficacy measures (88%). We converted the *p*-values into $Z$ values and estimated $\hat{g}(\theta)$ using Efron's deconvolution method [8]. For the rest of this section, all results are based on $\hat{g}(\theta)$ that was estimated on the assumption that the 12% trials we could not retrieve *p*-values on are all negative trials with $p \geq 0.05$. This assumption is conservative in the sense $\hat{g}(\theta)$ could be biased to the left if some of the 12% of trials are indeed positive trials.

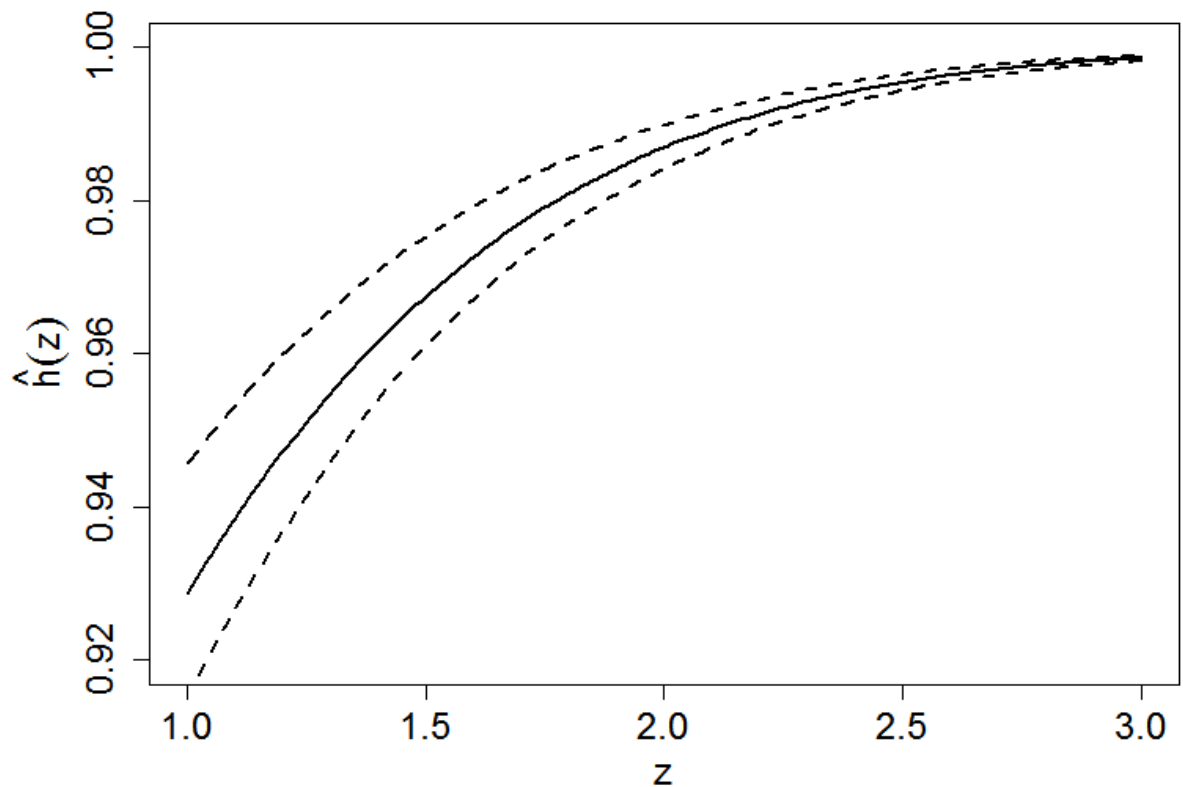

**Figure 1.** $\hat{h}(z)$ as a function of $z$ based on a conservative estimate of $\hat{g}(\theta)$ [5]. Dashed lines are 95% confidence intervals based on 500 bootstrap samples.

The estimated $\hat{g}(\theta)$ leads to a $\hat{p} = 9\%$ (95% CI: 6.8-11.1%) [8], suggesting that about one efficacy measure out of 10 efficacy measures of the included Phase III RCTs is null. Thus, in the setting of frequentist hypothesis testing where $m = 1$ for all trials, the total error is $\tau_0/\hat{p} \approx 11\tau_0$. If $\tau_0 = 1$ for a given period of time (e.g. no more than one expected number of false positive), then the total error is 11. With a fixed type I error rate of 2.5% for every trial, it means 440 trials should be allowed within the projected period of time. A non-uniform spending of the total error could lead to more or less number of trials. Fig 1 shows the $\hat{h}(z)$ as a function of $z$, where $\hat{h}(z)$ is the estimated $h$-probability based on $\hat{g}(\theta)$. In particular, the conventional threshold of $z = 1.96$ (corresponding to two-sided $p$-value of 0.05) is associated with a $h$-probability of 0.986 (95% CI: 0.983-0.989); threshold of $z = 1.65$ (corresponding to two-sided $p$-value of 0.1) is associated with a $h$-probability of 0.975 (95% CI: 0.970-0.981).

## 5 Discussions

In this article, we focus on two major limitations of current statistical inference of Phase III RCTs: (i) uniformly applied threshold on test statistics may not suit all scenarios; (ii) there is no existing method to control/estimate the number of false positives yielded from the large number of Phase III RCTs. We proposed two frameworks to address the two limitations based on the frequentist hypothesis testing paradigm and Bayesian perspectives. As an illustrative example, we estimated distributional parameters (e.g. $\hat{g}(\theta)$) for industry sponsored Phase III superiority RCTs that can be used to execute the proposed strategies. It should be noted that the Bayesian solution can also address issues (a)-(c) around the very concept of statistical evidence of the efficacy raised at the beginning of the article. Specifically, the posterior probability of non-null [6] and credible region of the actual efficacy measure [5] (not discussed in this article) are heuristically more accessible to investigators for a good understanding of uncertainty [2].

Key to the execution of the proposed frameworks is sufficient data for accurate $\hat{g}(\theta)$. As explained in the previous section, the initial FDAAA does not require trial results to be submitted to CT.gov for medical products not approved by FDA [13]. This has made results retrieval a challenging task for some of the 1,393 trials discussed in the previous section. A predominant amount of efforts was devoted to the search of trial results not available on CT.gov in our previous study. The US Department of Health and Human Services, under the designation of FDAAA to issue regulations regarding certain statutory provisions, has issued a new rule on September 16, 2016 [16]. Under this new rule, trials on essentially all medical products, regardless of FDA approval, are required to report summary results on CT.gov. Thus, the proposed strategies are readily feasible with the support of the FDAAA and its expanded rules. In particular, it is conceivable that $g(\theta)$ may not be static over time, particularly in consideration of trial sponsors' response to a policy that allows for more flexible error control. Therefore, $\hat{g}(\theta)$ can be regularly updated to incorporate results of most recent trials. The execution of such a regular update is realistic with the new rules that provide guarantee in data support.

The bound for ENFP based on the Frequentist framework is conservative (higher) as it assumes $\beta = c$ for a null efficacy trial even in fact $\beta < c$, e.g. chance of rejecting the null is higher for $\beta = c$ than for $\beta < c$. Bayesian framework is less conservative as it accounts for the actual distribution of the effect size and subsequently factors in the actual values of those $\beta$ values less than $c$. On the other hand, the Frequentist framework only relies on $\rho$, which is more stable than $g(\theta)$. For instance, once the policy of flexible false positive control is first implemented, trials sponsors will respond by changing their strategy on which medical products to be put on Phase III. In addition, the sample sizes will change based on the false positive error assigned to them. The first factor will affect $\rho$ but both factors will affect $g(\theta)$. Thus, at the start of the new policy, the $\hat{\rho}$ computed from historical trials is perhaps a more reliable estimate for the $\rho$ of the trials under the new policy as compared with $\hat{g}(\theta)$ as an estimate of $g(\theta)$. Therefore, the Frequentist framework is more suitable for the beginning of a new policy. Once the policy is in place for a while and the dynamics reaches an equilibrium, the Bayesian framework provides a more accurate error control mechanism.

A natural question following the development in this article is how to determine the rejection region for a trial if it is allowed to be different from trial to trial. As the concordance assumptions are the main requirement for our solutions, it is important to hold it true when designing the strategy for the determination of the rejection region. We project that the concordance assumptions will hold provided that the criteria used to determine the rejection region root in other evidence on the efficacy of the medical product being investigated such as earlier phase evidence of efficacy, biological plausibility, evidence from similar medical products etc. Even if regression to the mean may play a role here (e.g. strong efficacy signal from earlier phase may be due to a large error that occurred in thousands of compounds being investigated), it is extremely unlikely that it will render the concordant assumptions invalid. There are, however, other factors with unclear indication of efficacy of a medical product. For example, the existing treatments available to a specific condition, the severity of the disease and its impact on quality of life, the size of the patient population, and other factors could affect the perceived public health impact of a new medical product. Whether or not these factors should be factored into the equation on the determination of the rejection region and if yes how to implement it are certainly subject to debate. Statistically, these factors could negatively affect the concordant assumptions, e.g. relaxed rejection region for a new product due to lack of existing treatment can work against the concordant assumptions because the fact that there is no existing treatment could mean high probability that the new product has no efficacy.

Therefore, these factors should be studied and weighted carefully to determine their roles in the determination of the rejection region.

In summary, the frameworks proposed in this article have the advantages of flexible false positive control for individual trials and control of the overall number of false positives. It also is practically feasible because the distributional parameters required for its execution can be readily estimated and updated using data at CT.gov. As Phase III RCTs serve as the gold standard to evaluate the efficacy of a medical product, its direct impact on public health and advancement in biomedical research is profound. Statistical criteria are the driving force of the conclusion of an RCT, which speaks for its importance. There has been active debate in the statistical and scientific communities on the transition from the well-known rule of "$p<0.05$" to better strategies. Our solutions, in the same spirit, offer an effective solution in the setting of Phase III randomized trials. We hope to draw more attention on this topic to ultimately improve the way the data are interpreted for randomized clinical trials, both individually and collectively.

## Acknowledgement

We thank Prof. Bradley Efron from Standard University for providing computer programs for the analysis. We thank Prof. RongHui Xu from University of California at San Diego for constructive suggestions.

## Funding

None